\documentclass[preprint,
superscriptaddress,
amsmath,amssymb,aps,showkeys,showpacs,
twoside,final,secnumarabic,
nofootinbib]{revtex4-2}

\usepackage[paperwidth=205mm,paperheight=290mm,top=17mm,bottom=25mm,
inner=17mm,outer=17mm,
twoside]{geometry}

\usepackage{cmap} 
\usepackage[T1,T2A]{fontenc}
\usepackage[utf8]{inputenc}
\usepackage[russian,english]{babel}
\usepackage{color}
\usepackage{graphicx}
\usepackage{dcolumn}
\usepackage{bm} 
\usepackage[unicode=true,colorlinks=true,linkcolor=magenta, urlcolor=blue, citecolor = blue,breaklinks]{hyperref}
\usepackage{multirow}
\usepackage{url}
\usepackage{breakurl}
\DeclareGraphicsExtensions{.eps}

\newcount\issue
\newcount\Vol
\newcount\numb
\usepackage{fancyhdr} 
\def\Vol{\textbf{80}}
\def\numb{x}
\begin{document}

\title{Photometric light curves analysis of SU~UMa-type dwarf novae: \\the case of RZ~LMi and KV~Dra} 

\def\addressa{Sternberg Astronomical Institute, Moscow State University, Universitetsky pr., 13, Moscow 119234, Russia}
\def\addressb{Institute of Astronomy  of the Russian Academy of Sciences, 48 Pyatnitskaya st., Moscow 119017, Russia}

\author{\firstname{I.B.}~\surname{Voloshina}}
\email[E-mail: ]{voloshina.ira@gmail.com}
\affiliation{\addressa}
\author{\firstname{A.N.}~\surname{Tarasenkov}}
\affiliation{\addressb}
\affiliation{\addressa}

\received{xx.xx.2025}
\revised{xx.xx.2025}
\accepted{xx.xx.2025}

\begin{abstract}
The results of new photometric observations of two SU~UMa type dwarf novae are presented. Superhumps were detected on the light curves of these systems during superoutbursts and their periods and amplitudes determined. The classification of objects as dwarf novae of the ER~UMa subclass (for RZ~LMi) and subclass WZ~Sge (for KV~Dra) has been clarified. Orbital period $0^{d}.0586$ for KV Dra was determined.
\end{abstract}

\pacs{Suggested PACS}\par 
\keywords{Close binary systems, cataclysmic variables, dwarf novae, photometry, stars - individual: RZ~LMi, stars - individual: KV~Dra   \\[5pt]}

\maketitle
\thispagestyle{fancy}


\section{Introduction}\label{intro}

 SU~UMa type systems are subclass of dwarf novae with orbital periods less than 2.5 hours. They consist of a white dwarf primary and low mass main sequence star as a secondary one. The latest fills its Roche lobe and transfers matter to the white dwarf, forming an accretion disk around it. These objects usually show two kind of outbursts: normal outbursts and so-called superoutbursts which exceed the normal ones both in brightness and duration of the outburst. During superoutbursts they exhibit brightness increase on the small part of the light curve repeated with the period a few percents longer than the orbital one with amplitudes around $0.1- 0.3^m$. Superhumps are due to gravitational disturbances from secondary (tidal-thermal model~\cite{1}). These disturbances became most effective than the matter of accretion disk comes to 3:1 resonance with orbital motion of the secondary. The beating of the orbital and precessional periods cause periodic variations, identified as superhumps. 
 
 SU UMa type stars, in turn, are divided into two subclasses: WZ~Sge-type and ER~UMa-type. The first ones have the shortest orbital periods among dwarf novae (about 80-90 min), their outbursts occur less frequently and with a larger amplitude (about $8^m$). In such systems complex post-superoutburst rise of brightness (rebrightening or echo outbursts) could be observed. The absence of normal outbursts is also defining characteristic of such stars.

Dwarf novae of ER~UMa-type represent the small subgroup of SU~UMa stars with extremely short interval between superoutbursts (about 20-50 d) and normal outbursts occurring every 4 days.  

For our study we chose two dwarf novae belonged to the different subtypes of SU~UMa stars - RZ~LMi and KV~Dra.
 
\section{\label{sec:RZ}RZ~LMi}

RZ~LMi was originally discovered as a variable with ultraviolet-excess by Lipovetskii and Stepanyan~\cite{2}. Green et al. (1982) later confirmed it spectroscopically as a cataclysmic variable (CV)~\cite{3}. Robertson and his colleagues in 1995 performed photometry of RZ~LMi and two other stars with similar behavior. As a result they found similar characteristics of these stars. This allowed them to combine these objects into one group called type ER~UMa stars.

Olech et al. (2008)~\cite{4}, Nogami et al. (1995)~\cite{5} and other observers detected superhumps in the light curve of this system and obtained some statistical characteristics of RZ~LMi such as cycles for normal outbursts and superoutbursts, duration of normal outburst and superoutbursts, its magnitude at active and quiet state. The periodic variations connected with orbital period of the binary were never detected before for RZ~LMi. The orbital period $0^{d}.05792$ was estimated later by T. Kato et al. (2016)~\cite{6} from consideration of different facts, but not as result of direct observations.

\section{\label{sec:KV}KV~Dra}

The RX~J14505+6403 system was discovered during the ROSAT All-Sky Survey as a possible cataclysmic variable of unknown type with a magnitude of $16.3 - 17.1$~\cite{7}. The first mention of this system as a possible dwarf nova appeared in 2000~\cite{8} when the first outburst of this object was recorded in May 2000. During intensive observations, superhumps were detected in the light curves of KV Dra and their period was determined as  $0^{d}.0601{\pm}0.0002$ and an amplitude about $0^{m}.3$. Later, Thorstensen~\cite{9} obtained an orbital period value of $0^{d}.0588$ as a result of spectral observations in quiescence in April 2000. Nogami (2000)~\cite{10} observed the system by photometrically means at inactive state and determined the period of $0^{d}.056$, but with a large error. All these data evidenced that the object is a typical dwarf nova of the SU~UMa type. Subsequent outbursts in this system were registered by Japanese scientists in 2002, 2004 and 2005, but duration of the superoutburst cycle has never been precisely determined.

\section{\label{sec:Obs}Observations}

Our observations of dwarf nova RZ~LMi were carried during 20 nights from February - April, 2024 and prolonged on the next year, in March (10 nights), with  50–cm telescope of INASAN Kislovodsk optical station with sCMOS camera ZWO ASI6200MM Pro in V band. Observations covered 3 normal outbursts and 4 superoutbursts of  RZ~LMi. The accuracy of our data is about $0.^{m}02-0.^{m}03$ (inactive state) and $0.^{m}08-0.^{m}011$ (outburst). Description of our equipment is given in paper~\cite{11}.

In addition, unpublished observations of this star in April 2019, obtained with 50-cm and 60 cm telescopes of the Sternberg Astronomical Institute Crimean Observation Station in the V and Rc bands (3 nights), were also used for analysis.

Observations of KV~Dra were performed with the same equipment in March 2025 when the system was in inactive state. Therefore for this dwarf nova, we also used unpublished observational data obtained during superoutburst in 2009 on the 60 cm telescope in Crimea.

\section{\label{sec:LC}Light Curves}
\subsection{\label{sec:RZLC}RZ~LMi}
\begin{figure}[ht!]
\includegraphics[width=1\textwidth,clip=true,trim=0.0cm 0.0cm 0.0cm 0.0cm,angle=0]{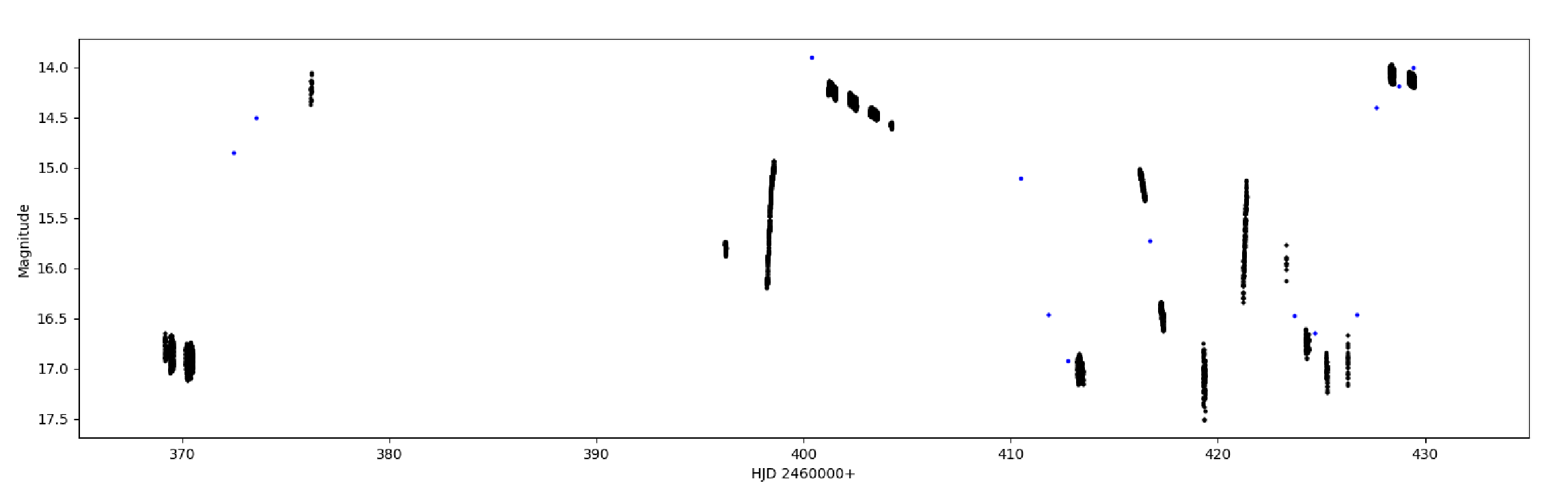}
\caption{The overall light curve of RZ LMi for the period of our observations in 2024.}
\label{fig:all}
\end{figure}

One of the two overall light curves of RZ~LMi obtained during our observations in 2024 and 2025 is shown in Fig~\ref{fig:all} as an example. There are three superoutbursts and two normal outbursts between subsequent superoutbursts. It is also evidence the very short duration of quiescence for this system.

Some of the daily light curves of RZ LMi obtained for different states of the outburst cycle are shown below in Fig~\ref{fig:RZ_d} as an example. Superhumps existed in the light curves during a superoutburst, but their amplitude and shape changes as the outburst develops. During the normal outburst there are no superhumps in the light curves of RZ LMi (see Fig.~\ref{fig:RZ_d}).

\begin{figure*}[ht!]
\centering
\includegraphics[width=0.44\textwidth,clip=true,trim=0.0cm 0.0cm 0.0cm 0.0cm,angle=0]{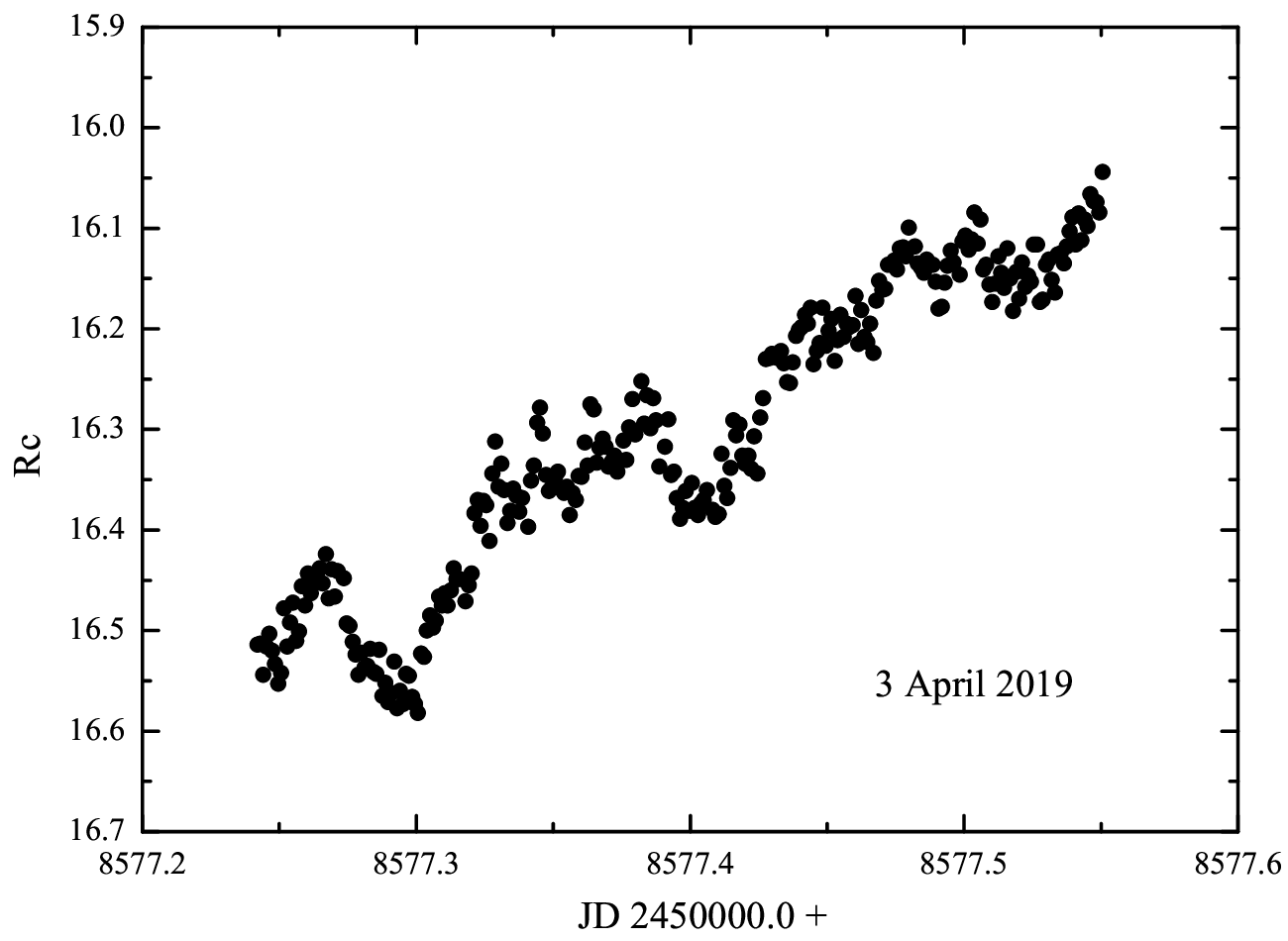}
\includegraphics[width=0.48\textwidth,clip=true,trim=0.0cm 0.0cm 0.0cm 0.0cm,angle=0]{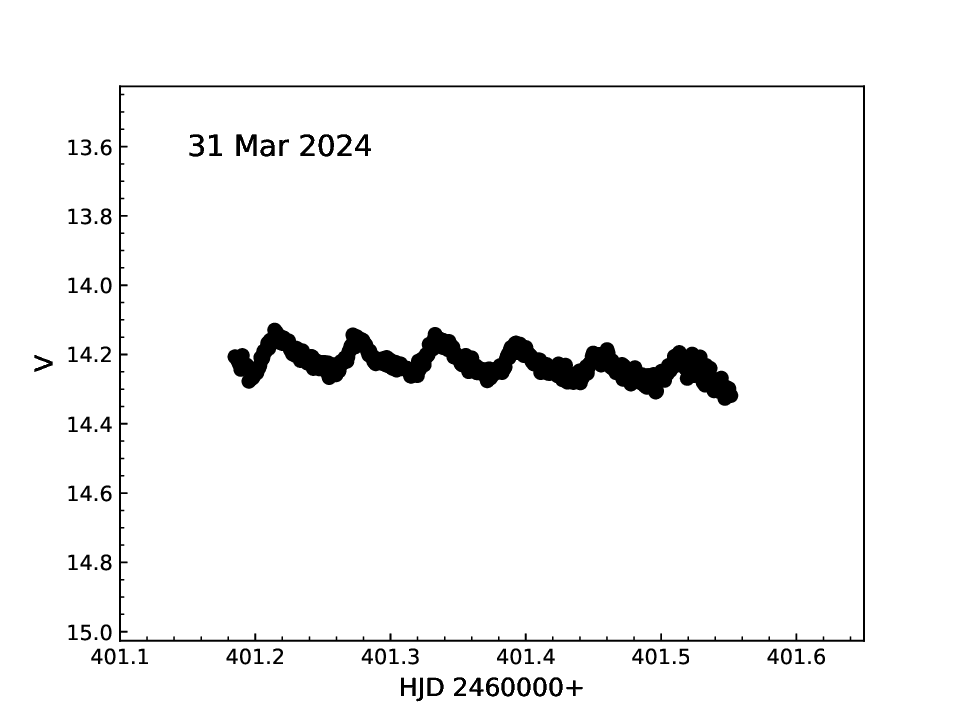}
\includegraphics[width=0.48\textwidth,clip=true,trim=0.0cm 0.0cm 0.0cm 0.0cm,angle=0]{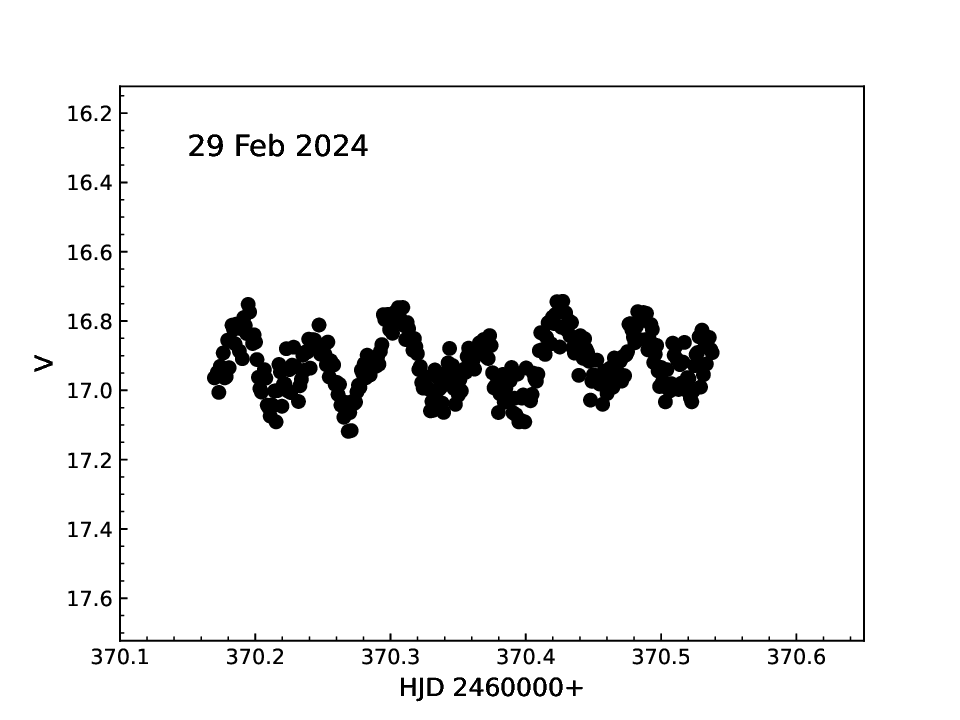}
\includegraphics[width=0.48\textwidth,clip=true,trim=0.0cm 0.0cm 0.0cm 0.0cm,angle=0]{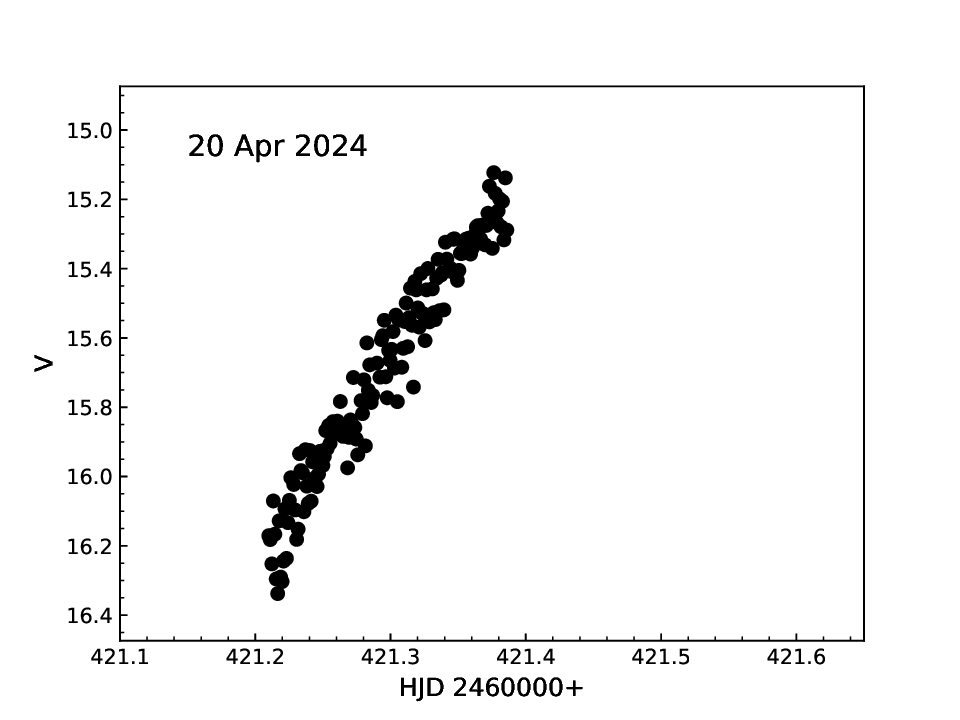}

\caption{Some of the individual light curves of RZ LMi: a). early superhumps during initial brightness rise at the beginning of superoutburst on April 3, 2019; b). well-developed superhumps in the light curve on March 31, 2024; c). inactive state of RZ LMi on Febrary 29, 2024; d) initial brightness rise at normal outburst – no superhumps in the light curve on April 20, 2024}
\label{fig:RZ_d}
\end{figure*}

\begin{figure*}[ht!]
\centering
\includegraphics[width=0.48\textwidth,clip=true,trim=0.0cm 0.0cm 0.0cm 0.0cm,angle=0]{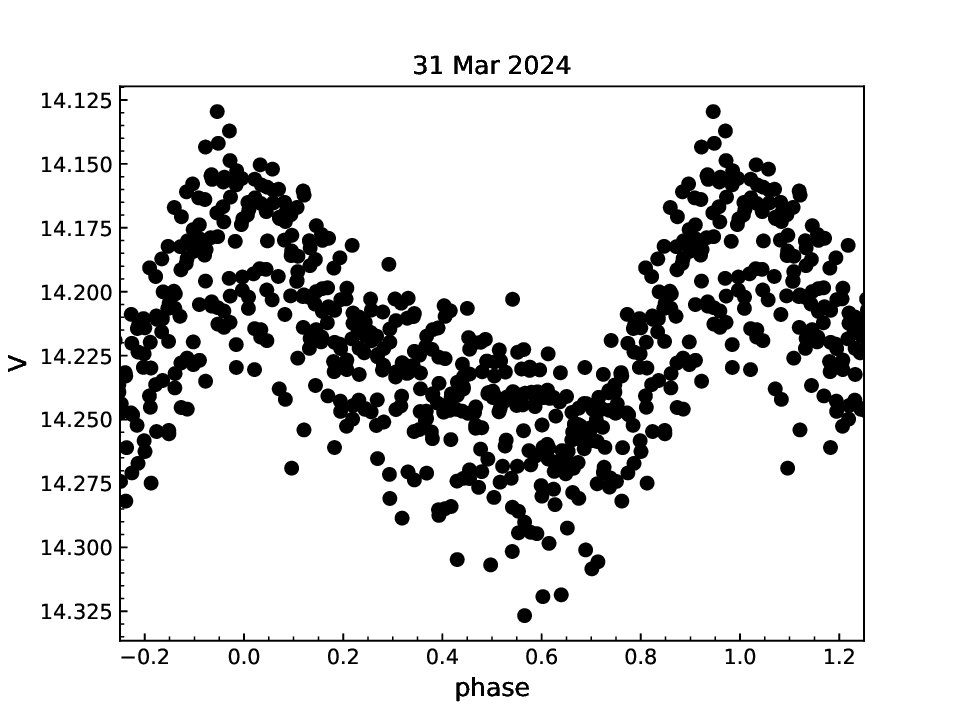}
\includegraphics[width=0.48\textwidth,clip=true,trim=0.0cm 0.0cm 0.0cm 0.0cm,angle=0]{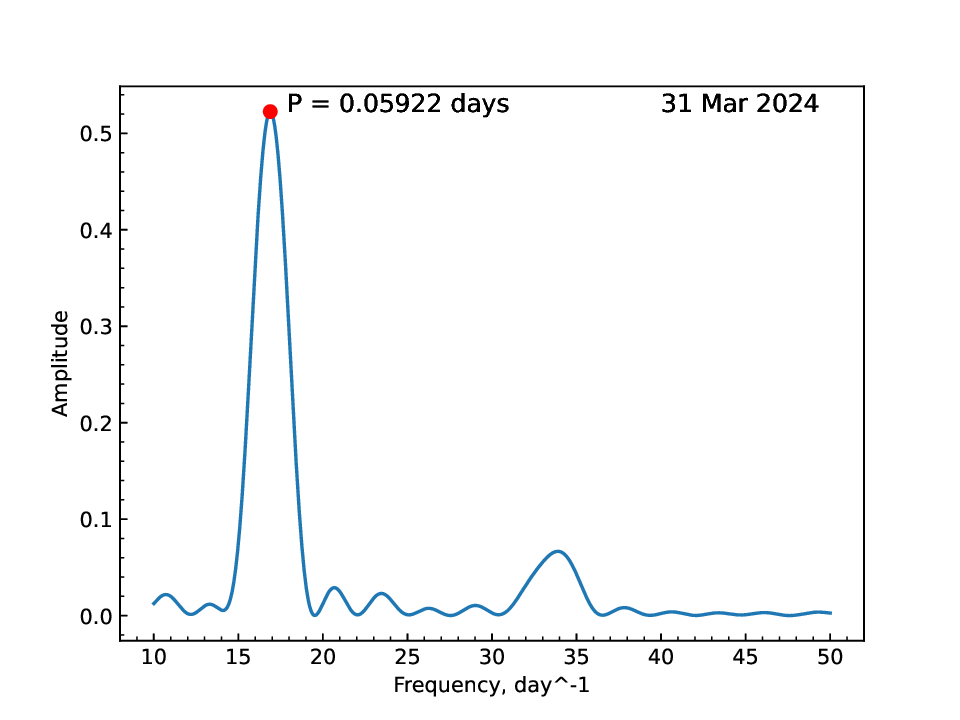}
\includegraphics[width=0.48\textwidth,clip=true,trim=0.0cm 0.0cm 0.0cm 0.0cm,angle=0]{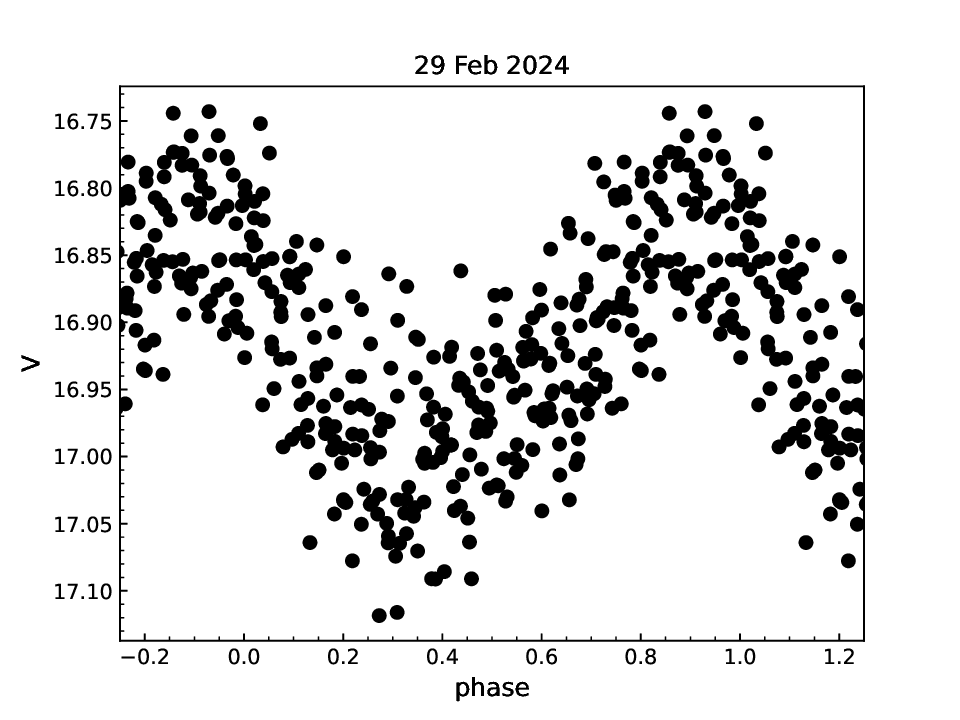}
\includegraphics[width=0.48\textwidth,clip=true,trim=0.0cm 0.0cm 0.0cm 0.0cm,angle=0]{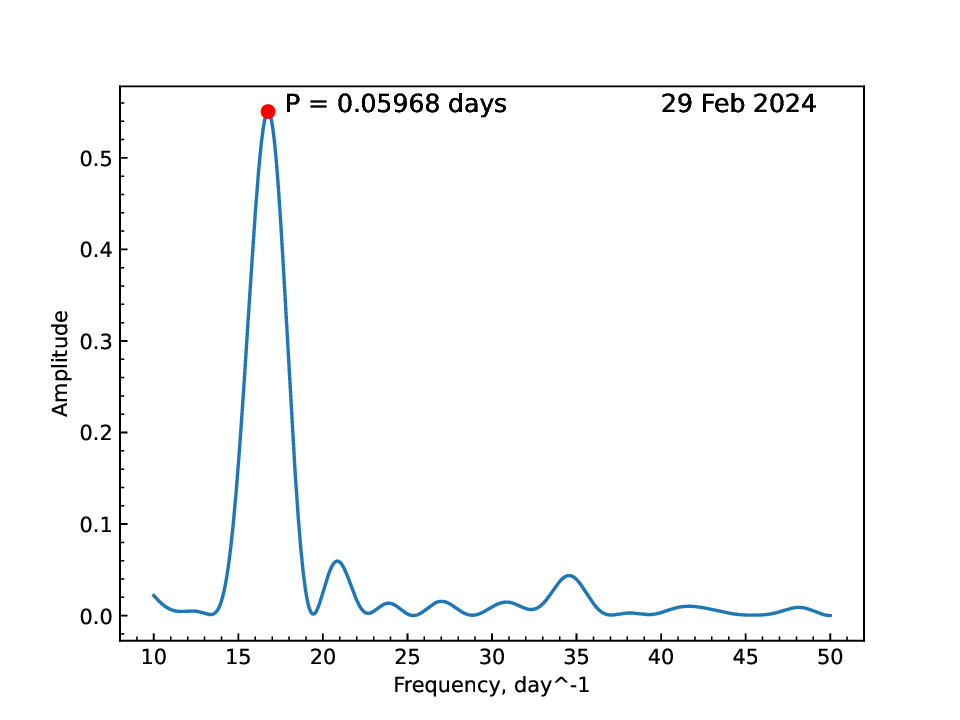}
\caption{Folded light curves (left) and periodograms (right) for RZ LMi. Upper panels: 31 March 2025 (superhumps), lower panels: 28 Febrary 2024 (inactive state). }
\label{fig:RZ_fold}
\end{figure*}

The search for periods was carried out using the Lomb-Scargle method~\cite{12}. Each observational set of superoutburst and quiescent state was analyzed separately. We searched for periods in the range of $0^{d}.02-0^{d}.10$. Accuracy of the obtained periods is about 0.0045 d. 
The power spectra and the corresponding light curves, folded with the obtained periods, are shown in Fig.~\ref{fig:RZ_fold}.

Despite the apparent similarity of the light curves in Fig.~\ref{fig:RZ_fold} during a superoutburst and in a quiet state (the presence of brightness variations), their properties and shape are completely different - in superoutburst the light curves are sawtooth, but in a quiet state they are different, bell-shaped and  the amplitude is also different. These may be short-period brightness variations observed in many dwarf novae at quiescence or probable orbital variations, however, due to the very short quiet state of the RZ LMi and insufficient observations in this state, the conclusion needs additional confirmation.

\subsection{\label{sec:KVLC}KV~Dra}

\begin{figure}[ht!]
\includegraphics[width=1\textwidth,clip=true,trim=0.0cm 0.0cm 0.0cm 0.0cm,angle=0]{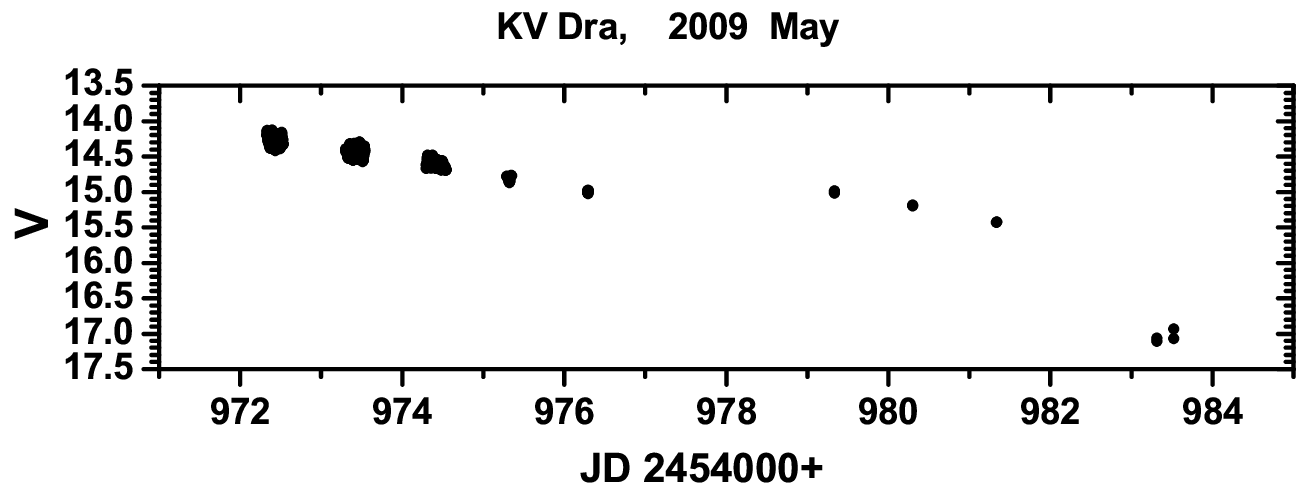}
\caption{The overall light curve of KV Dra for the period of our observations in 2009.}
\label{fig:KV_all}
\end{figure}

\begin{figure*}[ht!]
\centering
\includegraphics[width=0.51\textwidth,clip=true,trim=0.0cm 0.0cm 0.0cm 0.0cm,angle=0]{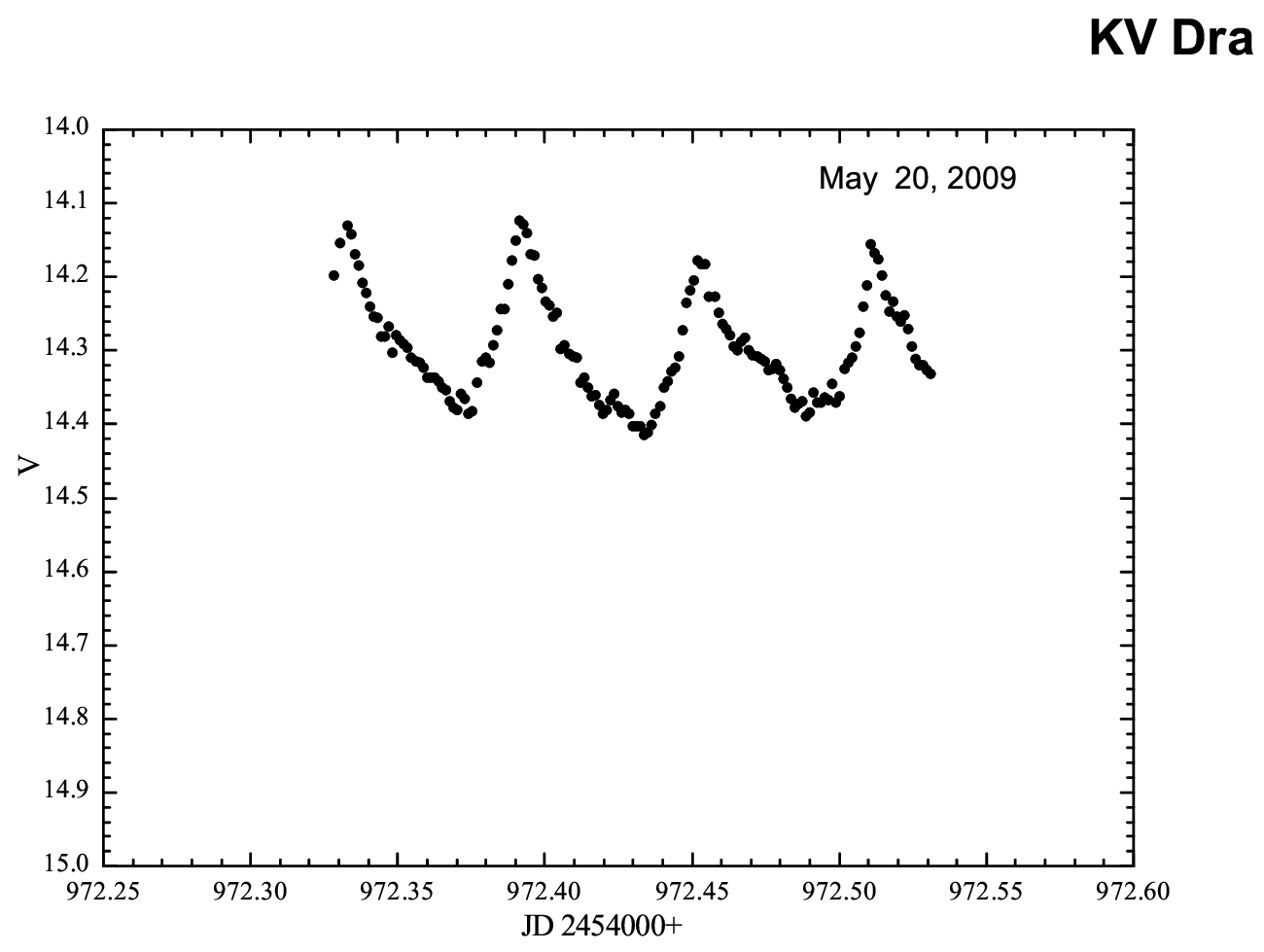}
\includegraphics[width=0.48\textwidth,clip=true,trim=0.0cm 0.0cm 0.0cm 0.0cm,angle=0]{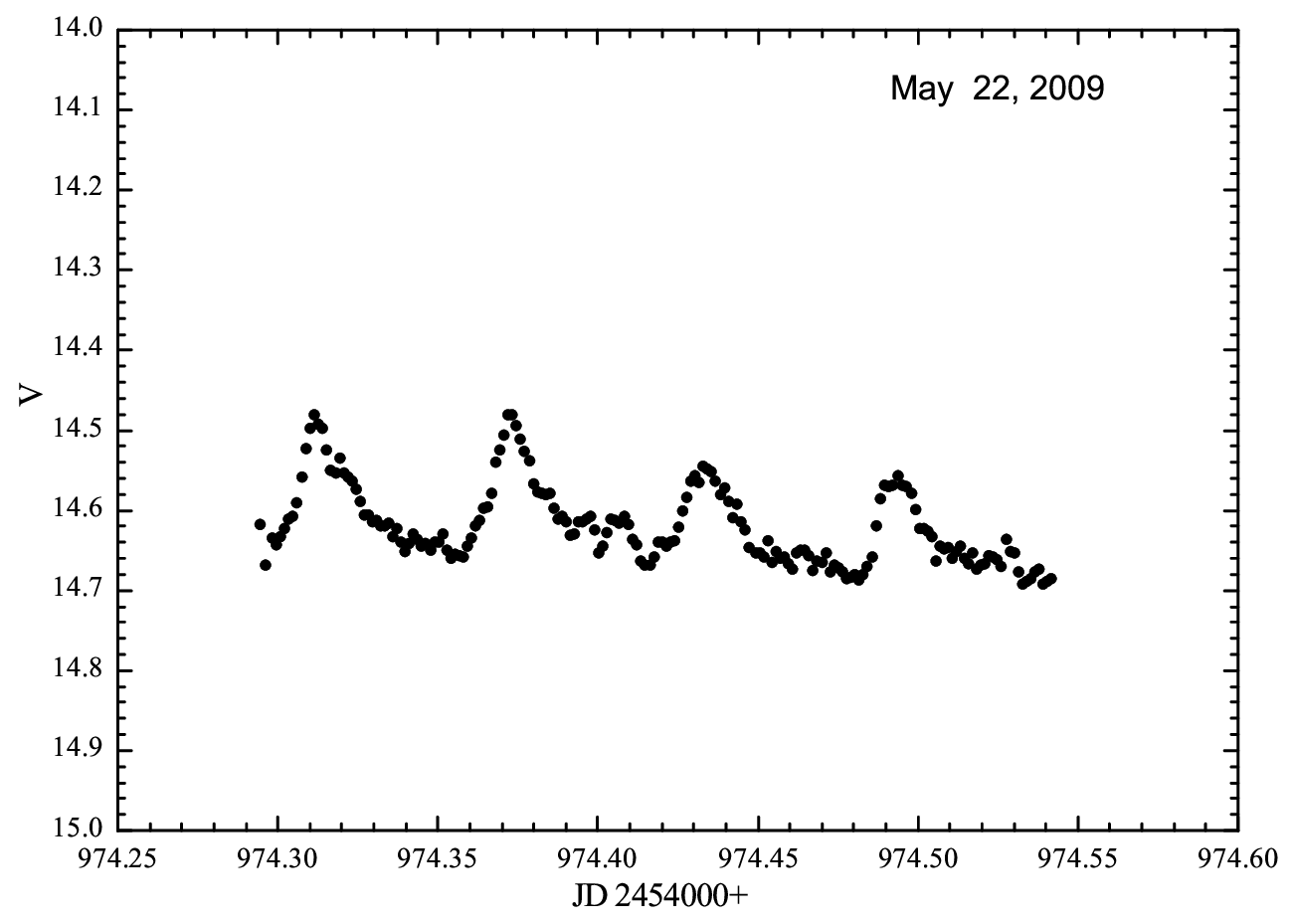}
\includegraphics[width=0.48\textwidth,clip=true,trim=0.0cm 0.0cm 0.0cm 0.0cm,angle=0]{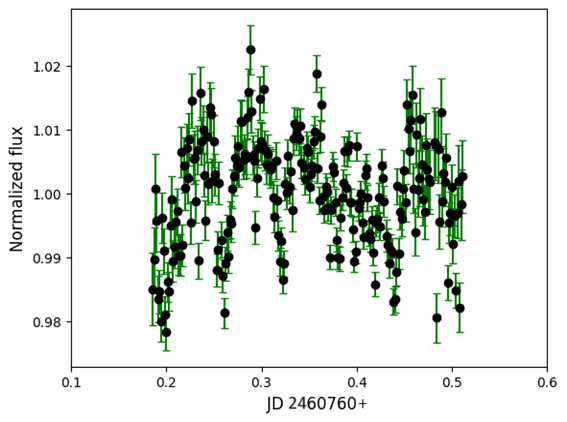}
\caption{Upper panels: daily light curves of KV Dra in superoutburst (20 and 22 May 2009), lower panel - KV Dra in quiescence (26 March 2025). }
\label{fig:KV_d}
\end{figure*}

The overall light curve of dwarf nova KV~Dra which was obtained during our observations in superoutburst in May 2009 is shown in Fig~\ref{fig:KV_all}. Below, in Fig~\ref{fig:KV_d} (a and b) one can see the individual light curves of this dwarf nova (for 20 and 22 May 2009). The light curve obtained on April 2025 during quiescence is demonstrated in lower panel of Fig~\ref{fig:KV_d}.

\begin{figure*}[ht!]
\centering
\includegraphics[width=0.48\textwidth,clip=true,trim=0.0cm 0.0cm 0.0cm 0.0cm,angle=0]{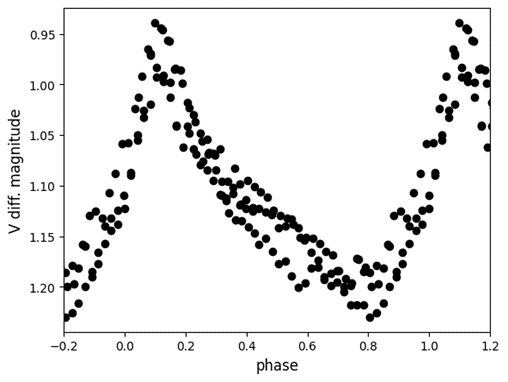}
\includegraphics[width=0.48\textwidth,clip=true,trim=0.0cm 0.0cm 0.0cm 0.0cm,angle=0]{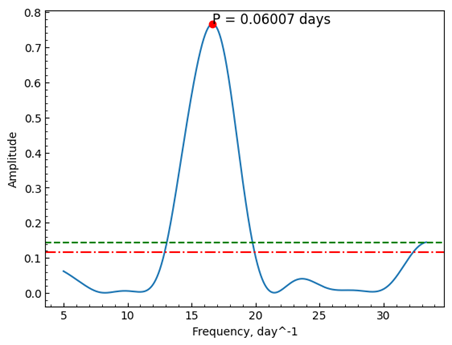}
\includegraphics[width=0.48\textwidth,clip=true,trim=0.0cm 0.0cm 0.0cm 0.0cm,angle=0]{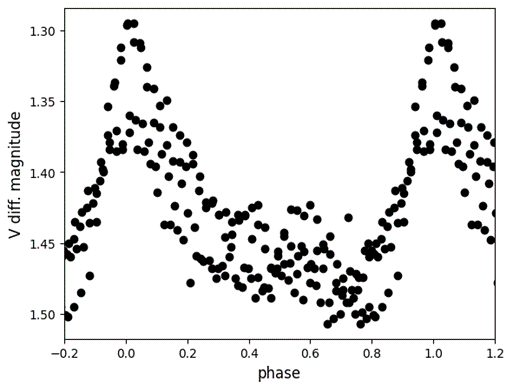}
\includegraphics[width=0.48\textwidth,clip=true,trim=0.0cm 0.0cm 0.0cm 0.0cm,angle=0]{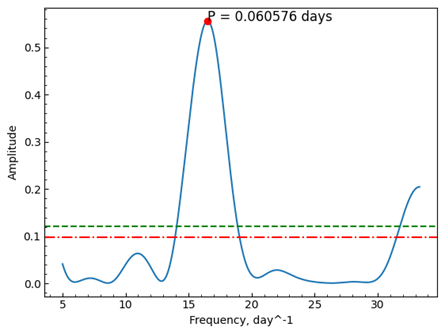}
\includegraphics[width=0.48\textwidth,clip=true,trim=0.0cm 0.0cm 0.0cm 0.0cm,angle=0]{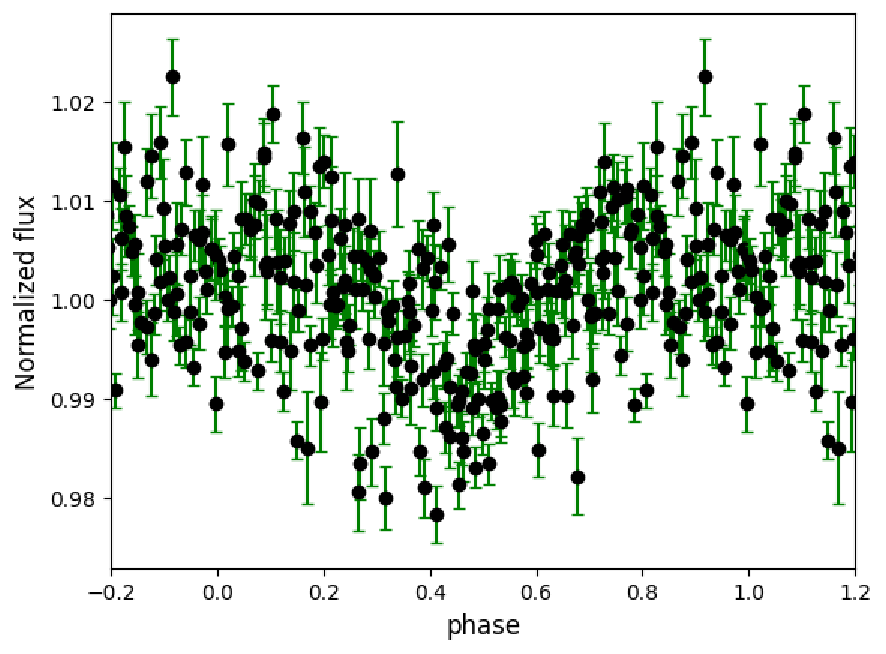}
\includegraphics[width=0.48\textwidth,clip=true,trim=0.0cm 0.0cm 0.0cm 0.0cm,angle=0]{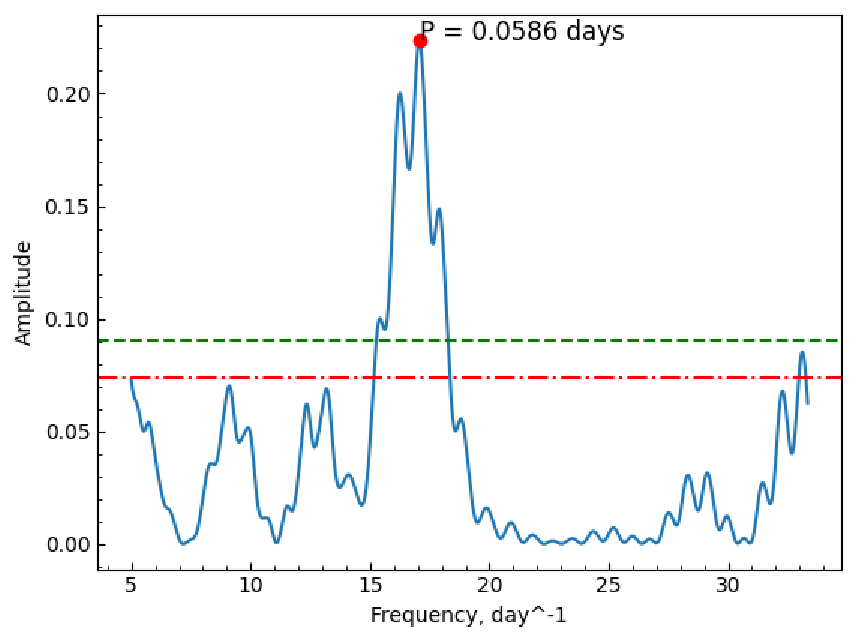}
\caption{Folded light curves (left) and periodograms (right) for KV Dra. Upper panels: 20 May 2009 (superhumps), medium panels: 22 May 2009 (superhumps), lower panels: 26 March 2025 (quiescence).}
\label{fig:KV_fold}
\end{figure*}

The power spectra were constructed during the determination of periods by the Lomb-Scargle method. Some of them together with the corresponding phase curves are shown in Fig.~\ref{fig:KV_fold} like an example.

\section{\label{sec:Conclusion}Conclusions}

The new photometric observations of two dwarf novae in active and quiet states
were made during 2024 and 2025. Based on analysis of these observations the light
curves of both stars were constructed. The following conclusions can be drawn
from consideration of these light curves:

For RZ LMi:

1) Superhumps were detected on the light curves of this system during
superoutbursts. They appear on the light curve 1.7 d after the
start of brightness rise and disappear after 7.2 d. At the beginning
they have a symmetrical sawtooth shape, later an asymmetry
occurs – the descending branch by the end of the visibility of the
superhumps significantly exceeds the ascending one in duration.

2) Duration of superoutburst is 13.8 d. The amplitude of the
superhumps at the beginning of the outburst exceeds $0^{m}.1$, and
then decreases to $0^{m}.06$. Duration of a normal outburst is $3^{d}.8$.

3) The average magnitude in V band is $14^{m}.2$ (superoutburst) and $16.^{m}85$ - $17^{m}.0$ (quiescence).

4) The light curve during superouburst have a characteristic saw-tooth shape with variation amplitude approximately $0^{m}.2$. The light curve outside of normal outburst is bell-shaped.

5) The superhumps detected in the light curves on April 28 and 29,
2024 (2-nd superoutburst in Fig 1) belong to stage A (growing
phase of superhumps according to classification given in paper~\cite{13}). Our value of superhumps periods for this stage are in
good agreement with results given in~\cite{13}. The superhumps periods
obtained for other dates are mainly those of stage B.

6) We have not been able to determine the orbital period of RZ~LMi,
since the quiet states in this system are very short, which also
makes it difficult to measure radial velocities. The almost
constant presence of superhumps on the light curves prevents the
detection of orbital changes in brightness by photometric
methods. Without knowledge of the orbital period and in the
absence of radial velocity measurements, it is impossible to
determine the mass ratio q directly from observations. Thus, the
evolutionary status of RZ LMi remains uncertain, despite the fact
that the outburst behavior and properties of the system are fairly
well known. Kato (2016)~\cite{6} concludes that the high value of mass
ratio q (even higher than for ordinary SU UMa dwarf novae) is
consistent with the rapid growth of superhumps in the system. He
suggested that RZ~LMi is a “bridge” between ER~UMa dwarf
novae and nova-like systems.

7) Our new observations of RZ~LMi give evidence that it is a member of
ER UMa dwarf nova.

For KV Dra:

1) Our photometric observations of KV~Dra during a superoutburst in May
2009 permitted to detect superhumps in the light curves of this system.

2) The periods of superhumps determined by these observations are the periods
of stage B, (intermediate stage, when the periods of superhumps varies~\cite{6}).

3) The value of the period obtained by observations in quiescence of $0^{d}.0586$
represented apparently the orbital period of the system. This value is in good
agreement with the spectral period determined by Thorstensen $0^{d}.0588$.
However, this conclusion requires confirmation from more numerous
observations with the larger telescope.

4) Our photometric observations of KV Dra during a superoutburst and in a
quiet state confirm the assumption that this is a typical WZ~Sge dwarf nova,
since it lacks normal outbursts, superoutbursts occur quite rarely, indicating a large outburst cycle and the amplitude of the outburst is quite high.

\begin{acknowledgments}
The study was conducted under the state assignment of Lomonosov Moscow State university.
\end{acknowledgments}

\section*{CONFLICT OF INTEREST}

The authors of this work declare that they have no conflicts of interest.


\end{document}